# Deterministic magnetization switching using lateral spin-orbit torque

*Yu Sheng[†], Yi Cao[†], Kevin William Edmonds, Yang Ji, Houzhi Zheng, and Kaiyou Wang[*]*

Y. Sheng, Dr. Y. Cao, Prof. Y. Ji, Prof. H. Zheng, Prof. K. Wang
State Key Laboratory for Superlattices and Microstructures
Institute of Semiconductors
Chinese Academy of Sciences
Beijing 100083, China
E-mail: kywang@semi.ac.cn

Dr. Y. Cao, Prof. K. Wang
Beijing Academy of Quantum Information Sciences
Beijing 100193, China

Prof. K. W. Edmonds
School of Physics and Astronomy
University of Nottingham
Nottingham NG7 2RD, United Kingdom

Prof. Y. Ji, Prof. H. Zheng, Prof. K. Wang
Center of Materials Science and Optoelectronic Engineering
University of Chinese Academy of Science
Beijing 100049, China

Prof. K. Wang
Center for Excellence in Topological Quantum Computation
University of Chinese Academy of Science
Beijing 100049, China

[†]These authors contributed equally to this work;
[*]Corresponding e-mail: kywang@semi.ac.cn

**Current-induced magnetization switching by spin-orbit torque (SOT) holds considerable promise for next generation ultralow-power memory and logic applications. In most cases, generation of spin-orbit torques has relied on an external injection of out-of-plane spin currents into the magnetic layer, while an external magnetic field along the electric current direction is generally required for realizing deterministic switching by SOT. Here, we report deterministic current-induced SOT full magnetization switching by lateral spin-orbit torque in zero external magnetic field. The Pt/Co/Pt magnetic structure was locally annealed by a laser track along the in-plane current direction, resulting in a lateral Pt gradient within the ferromagnetic layer, as confirmed by microstructure and chemical composition analysis. In zero magnetic field, the direction of the deterministic current-induced magnetization switching depends on the location of the laser track, but shows no dependence on the net polarization of external out-of-plane spin currents. From the behavior under external magnetic fields, we identify two independent mechanisms giving rise to SOT, i.e. the lateral Pt-Co asymmetry as well as out-of-plane injected spin currents, where the polarization and the magnitude of the SOT in the former case depends on the relative location and the laser power of the annealing track. Our results demonstrate an efficient field-free deterministic full magnetization**

**switching scheme, without requiring out-of-plane spin current injection or complex external stack structures.**

For more than a decade, the electrical switching of ferromagnets (FM) with perpendicular magnetic anisotropy (PMA), using spin-transfer torque (STT) and more recently spin-orbit torque (SOT), has underpinned the development of fast, low-power-consumption, and high-density spintronic devices[1-5]. In general, both the STT- and the SOT-induced switching of a FM layer require an injection of out-of-plane spin current from nearby layers[6-8]. For STT-induced FM switching, particularly, a spin-polarized current is generated in a magnetic tunneling junction (MTJ) structure when a charge current flows perpendicularly through the stacks, where another FM layer acts as a spin-polarizer[9]. Thus, device instability issues arise since the tunneling barrier layer between the two FM layers is required to transmit large switching currents.

The SOT-induced FM switching, on the other hand, circumvents this problem by using an in-plane switching current. Conventionally, a stack structure consisting of a strong spin-orbit coupling (SOC) layer and a FM layer is used, where an in-plane charge current gives rise to an out-of-plane pure spin current due to spin Hall effect (SHE) in the SOC layer and/or Rashba effect from the perpendicular interfacial inversion asymmetry[10-13]. The resulting SOT-induced effective magnetic field is in-plane, hence an additional orthogonal in-plane magnetic field is required to realize deterministic switching of a PMA-FM. To date, several approaches for field-free SOT-induced PMA-FM switching have been proposed and demonstrated, such as switching using a polarized ferroelectric substrate-induced in-plane spin current gradient[14-16], a wedge oxide capping layer[17], a tilted PMA layer[18], a stack with coherent in-plane exchange field[19-23], an interplay of SOT and STT[24,25], an in-plane-FM/normal metal/ PMA-FM trilayer[26], and a particular low symmetric $WTe_2$ semi-metal[27]. However, the concomitant complexities of these approaches highlight the inherent limitation of the conventional SOT scheme utilizing external out-of-plane spin current injection in a perpendicular asymmetric structure.

Here we demonstrate magnetic field-free deterministic current-induced magnetization switching in a PMA Pt/Co/Pt trilayer subjected to local laser annealing. Without external magnetic field, the direction of current-induced magnetization switching is found to depend on the relative location and the laser power of the annealing track, but is independent of the net spin current orientation from the two Pt layers. We attribute the observed behavior to a new SOT-induced perpendicular effective magnetic field originating from a lateral Pt gradient inside the Co layer. These results add further understanding to the physics of SOT and suggest a new scheme for magnetic field-free deterministic current-induced switching of a PMA-FM with more simplified stacks, even in the absence of out-of-plane spin current injections from neighbouring strong SOC layers.

**Engineering current-induced FM switching by laser annealing**

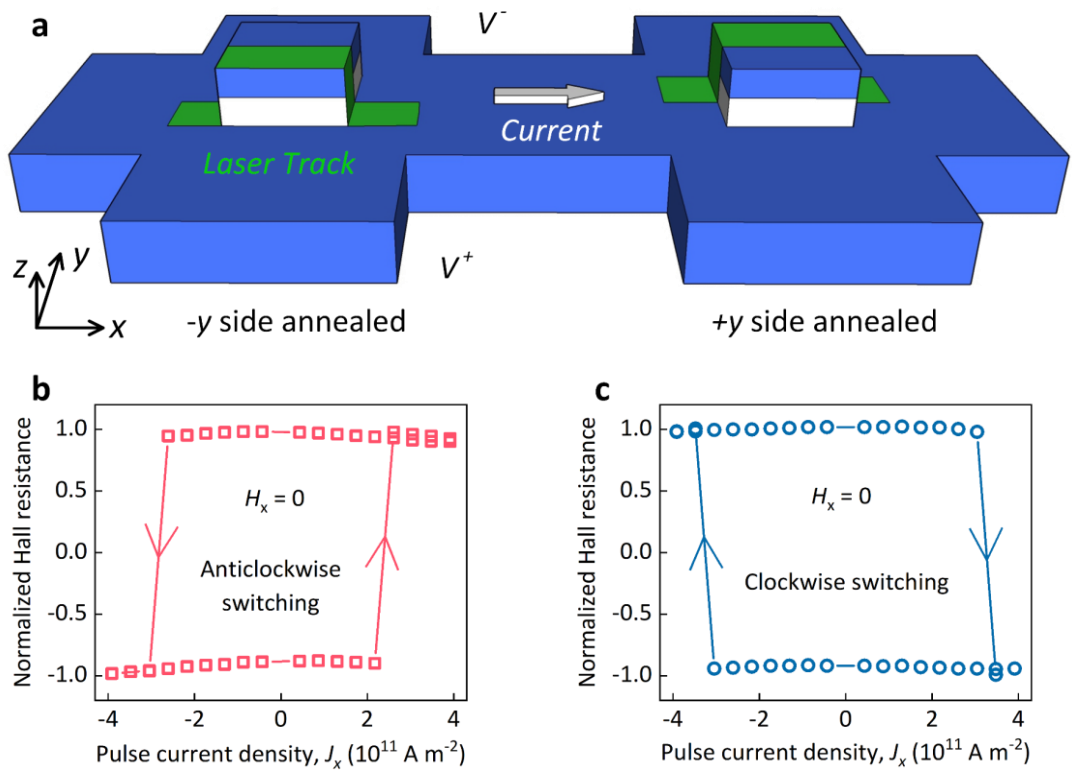

**Figure 1 | Schematic drawing of the locally laser annealed PMA-FM pillars and the respective current-induced magnetization switching. a**, Schematic of the Hall device with a stack structure of Pt(3)/Co(0.5)/Pt(2), where the top Co(0.5)/Pt(2) layers were fabricated into square pillars at each Hall cross. The Hall bar channel width and the pillar side length are 6μm and 2μm, respectively. Black arrows indicate the Cartesian (*x*, *y*, *z*) coordinate systems. Green zones illustrate the localized annealing tracks produced by an 8 mW laser, where the left/right device is annealed at the -*y*/+*y* side of the square pillar. **b-c**, In-plane (*x*-direction) pulse current induced perpendicular magnetization switching (represented by the normalized $R_H$-$J_x$ loops) without external magnetic field for (**b**) the -*y* side annealed pillar and (**c**) the +*y* side annealed pillar, respectively. The current

density $J_x$ was obtained by dividing the current intensity $I_x$ by the total cross-sectional area of both the 6μm-wide Pt(3 nm) Hall channel and the 2μm-wide Co(0.5 nm)/Pt(2 nm) pillar. The duration of every pulse is 10 ms and the anomalous Hall resistance $R_H$ was obtained by measuring the y-direction voltage under a small x-direction d.c. current (100 μA) 1 second after each pulse. To clarify the switching probability, the presented $R_H$ data was normalized by the out-of-plane magnetic field ($H_z$)-induced maximum Hall resistance of each pillar.

All devices discussed in this article showed good PMA with similar magnetic anisotropy field ($H_k$) both before and after the localized laser annealing, as illustrated in Supplementary S1. The schematic drawing of a locally laser annealed device is shown in Figure 1a. Stacks of Pt(3 nm)/Co(0.5 nm)/Pt(2 nm) (from the substrate side) were deposited on Si/SiO$_2$ substrates by magnetron sputtering. After deposition, the stacks were processed into 6μm-wide Pt(3 nm) Hall bars with 2μm-wide Co(0.5 nm)/Pt(2 nm) square pillars on top of each pair of Hall contacts. A laser with wavelength of 532 nm and power of 8 mW was used to locally anneal each pillar by sweeping across it along the x-direction, leaving a localized laser annealing track on the -y/+y side of the left/right pillar. A sequence of 10 ms current pulses was applied with a varying current density $J_x$, where $J_x$ was calculated by dividing the current intensity by the cross-sectional area of both the 6μm-wide Pt(3 nm) Hall channel and the 2μm-wide Co(0.5

nm)/Pt(2 nm) pillar. The perpendicular magnetization of each pillar was characterized by measuring the anomalous Hall resistance ($R_H$) using a small current of 100 μA following a 1 s interval after every pulse. As shown in Figure 1b and 1c, both locally laser annealed pillars showed 100% deterministic in-plane current-induced magnetization switching in the absence of external magnetic field, where the switching probability was given by the ratio of current-induced $R_H$ to the maximum external magnetic field induced anomalous Hall resistance $R_H^{max}$ (i.e. $R_H/R_H^{max}$) induced by an external out-of-plane magnetic field. Particularly, the sense of rotation of the $J_x$-induced switching loop is opposite between the -$y$ side (anticlockwise switching) and the +$y$ side annealed pillars (clockwise switching). This observed dependence of the switching sense on the location of the laser track could be exploited in integrated circuits with complementary spin logic and/or memory units[28,29].

**Zero-field FM switching without out-of-plane spin current injection**

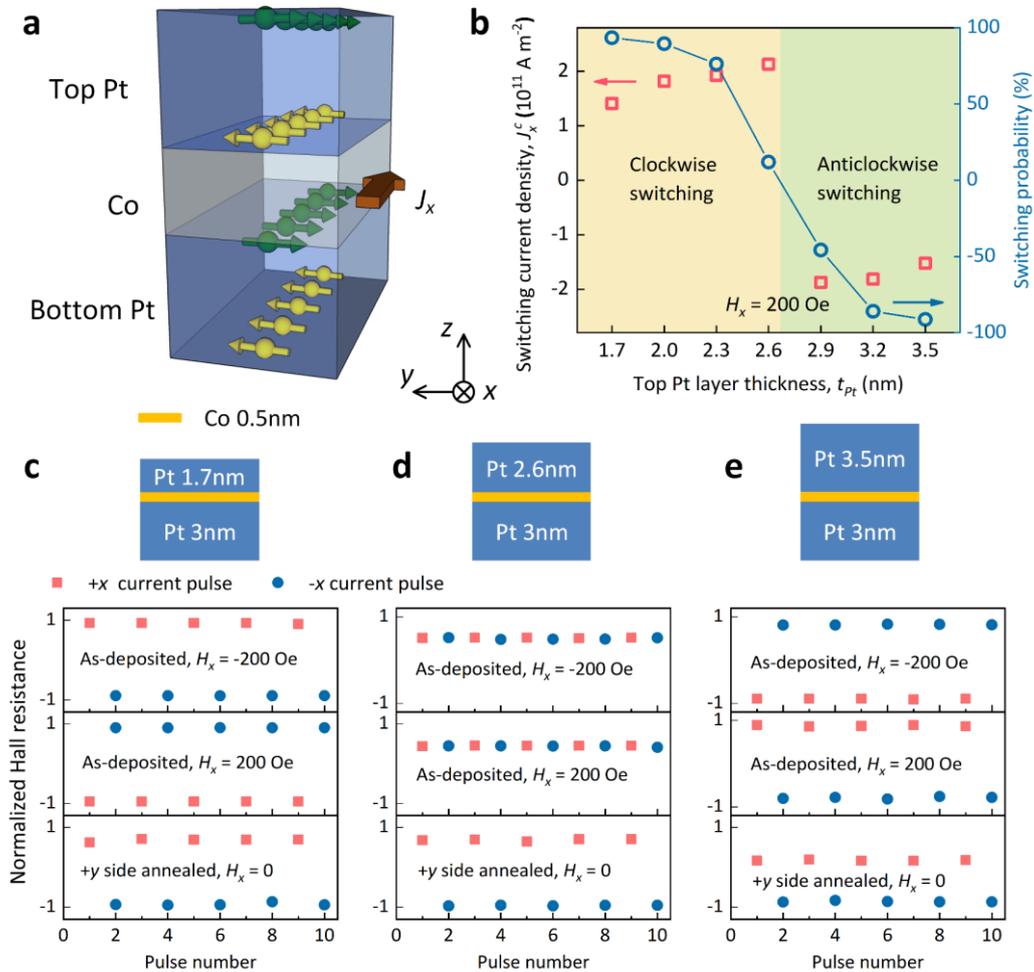

**Figure 2 | Current-induced magnetization switching of as-deposited and locally laser annealed Pt(3 nm)/Co(0.5 nm)/Pt($t_{Pt}$) trilayers. a**, Schematic of spin current injection for positive $J_x$ in a Pt/Co/Pt trilayer structure. The green and the yellow arrows represent spins polarized in the -$y$ and the +$y$-directions, respectively. **b**, Critical switching current density ($J_x^c$, red squares, defined as the average of the values of $J_x$ for magnetization switching from +$z$ to -$z$ directions) and switching probability (blue circles) versus $t_{Pt}$ for as-deposited Pt(3 nm)/Co(0.5 nm)/Pt($t_{Pt}$) samples. During the measurements, an in-plane magnetic field $H_x$ = 200 Oe was applied.

Positive/negative $J_x^c$ and switching probability denotes clockwise/anticlockwise switching, respectively. **c-e**, Magnetization switching with alternative positive and negative in-plane currents, for the as-deposited ($|J_x| = 2 \times 10^{11}$ A m$^{-2}$, $H_x = \pm 200$ Oe) and the +y side 16 mW locally laser annealed ($|J_x| = 2.8 \times 10^{11}$ A m$^{-2}$, without external magnetic field) samples with stack structures of (**c**) Pt(3 nm)/Co(0.5 nm)/Pt(1.7 nm), (**d**) Pt(3 nm)/Co(0.5 nm)/Pt(2.6 nm), and (**e**) Pt(3 nm)/Co(0.5 nm)/Pt(3.5 nm), respectively.

A series of Pt(3 nm)/Co(0.5 nm)/Pt($t_{Pt}$) trilayers were deposited and fabricated into Hall bars (without pillars) to investigate the dependence of the magnetization switching on the out-of-plane spin current for both as-deposited samples and locally laser annealed samples. As illustrated in Figure 2a, spin currents with opposite polarization are injected into the Co layer from the bottom and the top Pt layers. The threshold for current-induced magnetization switching was determined in an in-plane external magnetic field $H_x = 200$ Oe. The critical switching current density $J_x^c$ and the switching probability as functions of the top Pt layer thickness $t_{Pt}$ for the as-deposited samples are shown in Figure 2b, where we defined the positive/negative $J_x^c$ and switching probability ($R_H/R_H^{max}$) as representing the clockwise/anticlockwise switching respectively. The sense of rotation of the deterministic switching gradually evolves from clockwise to anticlockwise with growing $t_{Pt}$, indicating the injected spin

10 / 39

current is dominated by the bottom Pt layer for $t_{Pt} \leq 2.6$ nm and by the top Pt layer for $t_{Pt} \geq 2.9$ nm. For the approximately symmetric structure with $t_{Pt} = 2.6$ nm, the sample exhibits weak current-induced magnetization switching with a large $J_x^c$ of $2.13 \times 10^{11}$ A m$^{-2}$ and a small switching probability of 11.9%. This indicates that the spin currents from the bottom and the top Pt layers are mostly neutralized.

A comparison of current pulse-induced magnetization switching behavior between the as-deposited and the locally laser annealed samples is shown in Figure 2c-e. For $J_x = \pm 2 \times 10^{11}$ A m$^{-2}$ and $H_x = \pm 200$ Oe, no noticeable change in $R_H$ was observed in the as-deposited Pt(3 nm)/Co(0.5 nm)/Pt(2.6 nm) sample while binary $R_H$ with opposite signs were obtained in the as-deposited Pt(3 nm)/Co(0.5 nm)/Pt(1.7 nm) and Pt(3 nm)/Co(0.5 nm)/Pt(3.5 nm) samples. The sign of $R_H$ depends on all the three aspects, namely, the sign of $J_x$, the sign of $H_x$, and the stack structure (i.e. $t_{Pt} = 1.7$ nm in Figure 2c, or $t_{Pt} = 3.5$ nm in Figure 2e). However, the samples with $t_{Pt} = 1.7$ nm, 2.6 nm and 3.5 nm subjected to 16 mW localized laser annealing at the +$y$ side of their Hall cross areas (drawn schematically in Supplementary S1a) all showed substantial zero-field current-induced magnetization switching with the same switching sense regardless of the overall spin current polarization from the two Pt layers: the sign of $R_H$ depends only on the sign of the electrical current (here $J_x = \pm 2.8 \times 10^{11}$ A m$^{-2}$) and the location of the laser track. These results suggest the existence of a new, efficient, in-plane current-induced SOT with perpendicular effective magnetic field originating from the localized laser annealing, the direction of which is independent with the polarization of the injected spin current. In other words, current-induced magnetization switching without a net out-of-plane spin current injection can be realized in these locally laser annealed systems.

Furthermore, current-induced in-plane effective fields, which can be generated from an out-of-plane spin current injection, were also considered by conducting harmonic Hall measurements[12]. By analyzing the first and second harmonic signals of Pt(3 nm)/Co(0.5 nm)/Pt(2.6 nm) Hall bar samples (without pillar), as illustrated in Supplementary S2, no distinguishable damping-like torque can be determined for both the as-deposited and the 16 mW locally laser annealed samples. Out-of-plane spin current injections are thus once again shown, for both the as-deposited and the locally laser annealed sample, to be not strong enough to result in the current-induced magnetization switching. It further confirms that the current induced deterministic magnetization switching is coming from the localized laser annealing.

**Spin-orbit torque due to lateral Pt-Co asymmetry**

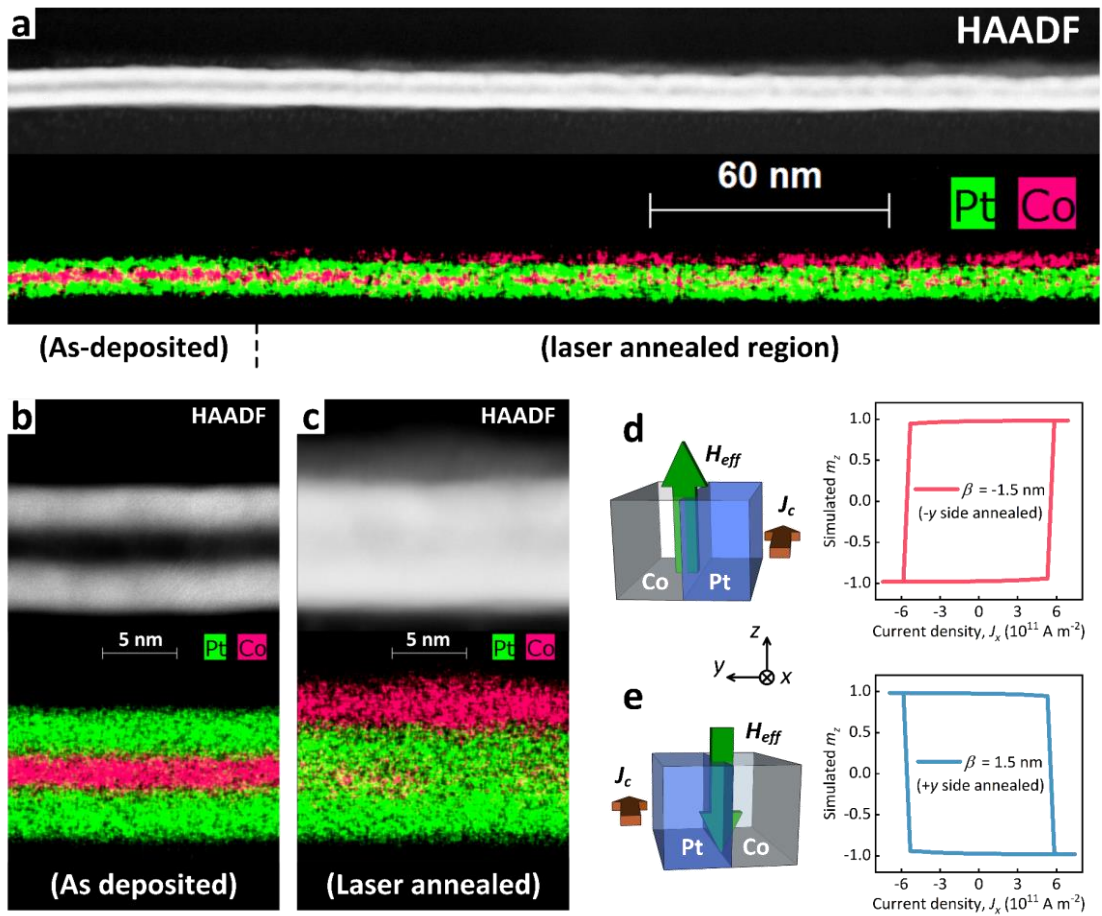

**Figure 3 | Cross-sectional STEM observation of the locally laser annealed Pt/Co/Pt trilayer and micromagnetic simulation of the current-induced magnetization switching in lateral Pt-Co asymmetric thin films.** Cross-sectional HAADF Z-contrast STEM images and corresponding EDS maps of a 12 mW locally laser annealed Pt(3 nm)/Co(3 nm)/Pt(3 nm) trilayer for (a) gradient area, (b) as-deposited region, and (c) laser annealed region, respectively. From bottom to top, there are the $SiO_2$ substrate surface, the Pt/Co/Pt trilayer, and the C capping layer. Brighter contrast corresponds to the heavier element (Pt) and darker contrast to the lighter elements (Co, Si, C). **d-e**, Simulated $M_z$-$J_x$ loops with localized laser

annealing at (**d**) the -*y* side (Pt/Co interface in +*y*-direction, given *β* = -1.5 nm) and (**e**) the +*y* side (Pt/Co interface in -*y*-direction, given *β* = 1.5 nm), respectively. The micromagnetic simulation was performed using Mumax3.

In order to reveal the nature of the deterministic current-induced magnetization switching in the locally laser annealed Pt/Co/Pt devices, we acquired microscopic pictures of both the as-deposited and the laser annealed regions. A cross-sectional STEM sample was fabricated from a 12 mW locally laser annealed Pt(3 nm)/Co(3 nm)/Pt(3 nm) 6μm-wide Hall bar using a focused ion beam (FIB) instrument, followed by high-angle annular dark field (HAADF) STEM, bright field (BF) STEM, and X-ray energy-dispersive spectrometry (EDS) experiments. The HAADF image contrast scales with the square of the atomic number (*Z*), i.e. brighter contrast represents heavier element (Pt) and darker contrast represents lighter element (Co, Si, C). On the contrary, the BF images give complementary, inverse contrast, as shown in Supplementary S3. Figure 3a shows the HAADF Z-contrast image and the corresponding EDS mapping of the gradient area containing the as-deposited and the laser annealed regions. The two regions are shown on an expanded scale in Figure 3b and 3c respectively. Distinct Pt/Co/Pt layers can be observed in the as-deposited region, while the laser annealed region shows significant intermigration between the Co and the top Pt layers, resulting in a Pt/Co($t_1$)/Pt/Co($t_2$)-like structure with laterally varying Co components $t_1$ and $t_2$. Extended HAADF and BF images covering the whole laser annealed region are shown in Supplementary S3, confirming that the effective width of the 12 mW laser annealing track is around 660 nm. The average laser power per unit area in this case (laser power = 12 mW) is then estimated as about $2.75 \times 10^{10}$ W m$^{-2}$. Since the intensity of a laser spot has a Gaussian distribution, there is no distinct boundary between the as-deposited and the laser annealed regions, and the ratio of $t_2$ to $t_1$ increases from the edge to the center of the laser annealed region. For a typical position near the center part of the laser annealed region, as shown in Figure 3c, most of the Co has migrated to the surface above the top Pt layers and is naturally oxidized into nonferromagnetic CoO$_x$[30,31].

The chemical composition gradient of Pt and Co elements results in a lateral interface between the laser annealed (mostly Pt) region and the as-deposited (Co) region. Analogous to the conventional SOT in perpendicularly asymmetric Pt/Co bilayers, the in-plane current $J_x$ induces a new SOT acting on the as-deposited Co region due to the lateral Pt/Co interface, with a perpendicular effective field given by

$$H_z^{eff} = \beta \hat{y} \times J_x \qquad (1).$$

The $\beta$ coefficient of this SOT relates the magnitude of the effective field with the degree of the lateral Pt-Co asymmetry, which depends on the laser power. The sign of $\beta$ depends on the relative location of the laser track, given as positive if the track is on the $+y$ side of the device. As shown in Figure 3d and 3e, micromagnetic simulations of the $m_z$-$J_x$ loops using an uniform magnetic film with given $β$ = -/+ 1.5 nm were performed by MuMax3 using Equation (1),[32] the switching senses of which are in accordance with the experimental results in Figure 1b and 1c, respectively.

# Dependence of the current-induced $H_z^{eff}$ on the laser power

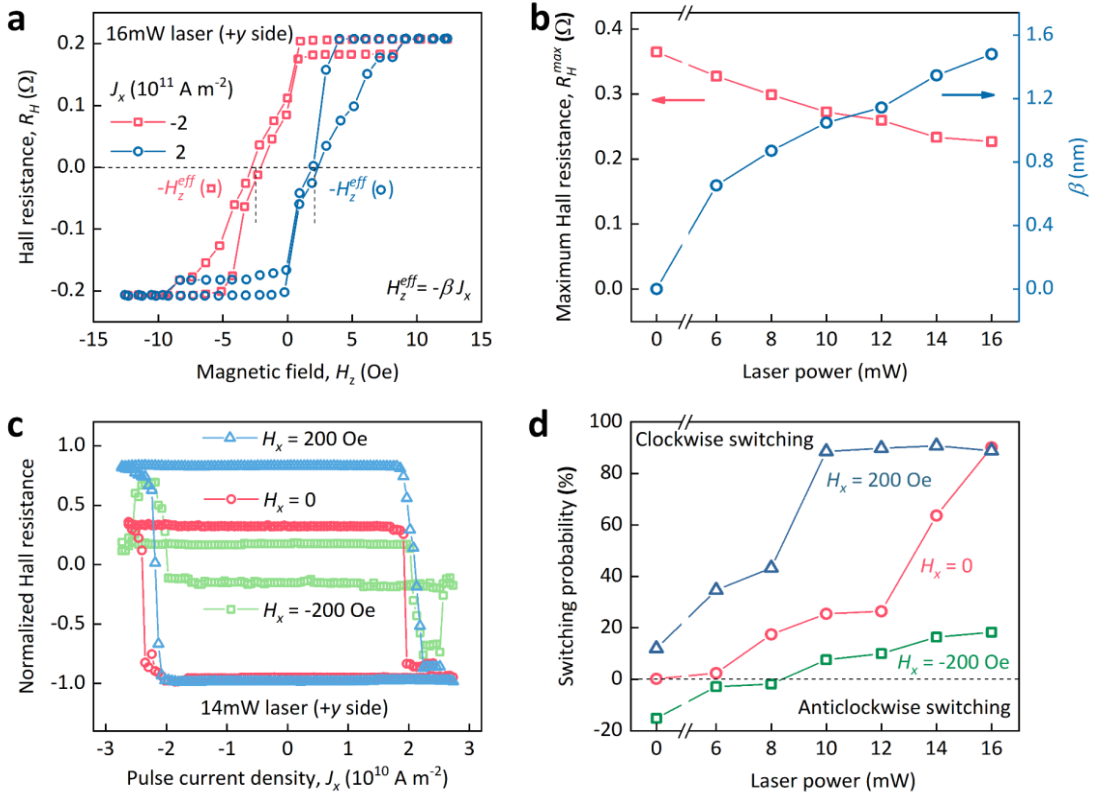

**Figure 4 | Current-induced effective magnetic field and magnetization switching for locally laser annealed Pt(3 nm)/Co(0.5 nm)/Pt(2.6 nm) Hall bars with varying laser power. a**, $R_H$-$H_z$ loops of the 16 mW +y side locally laser annealed sample under constant d.c. currents of $J_x = \pm 2 \times 10^{11}$ A m$^{-2}$. The $J_x$-induced effective magnetic field $H_z^{eff}$ was determined from each loop as the negative of its average coercivity field. **b**, The maximum Hall resistance $R_H^{max}$ obtained in an external out-of-plane magnetic field (red squares) and the SOT coefficient $β$ due to lateral Pt-Co asymmetry (defined as $β = -H_z^{eff}/J_x$, blue circles) versus laser power. **c**, Pulse current-induced $R_H$-$J_x$ loops of the 14 mW +y side locally laser annealed sample with $H_x = -200$ Oe (green squares), $H_x = 0$ (red circles), and $H_x = 200$ Oe

(blue triangles), respectively. **d**, The current-induced magnetization switching probability as a function of the laser power with $H_x$ = -200 Oe (green squares), $H_x$ = 0 (red circles), and $H_x$ = 200 Oe (blue triangles), respectively. The switching probability was obtained by dividing the $J_x$-induced maximum $R_H$ by the external field-induced maximum $R_H^{max}$, where a positive/negative switching probability denotes a clockwise/anticlockwise switching.

To quantitatively determine $\beta$ and further investigate its effect on the current-induced magnetization switching, $R_H$ loops were measured versus external out-of-plane field under fixed *dc* currents, for Pt(3 nm)/Co(0.5 nm)/Pt(2.6 nm) samples with local laser annealing on the +*y* side. As shown in Figure 4a, the $H_z^{eff}$ induced by a given $J_x$ can be extracted from the $R_H$-$H_z$ loop as its horizonal shift from zero point, where a left offset gives positive $H_z^{eff}$ while a right shift gives negative $H_z^{eff}$. The $\beta$ can thus be determined from $\beta$ = -$H_z^{eff}$/$J_x$, a scalar form of Equation (1). The increase of $\beta$ and reduction of maximum Hall resistance ($R_H^{max}$) with increasing laser power, as shown in Figure 4b and Supplementary S1, is in accordance with an increasing lateral Pt-Co asymmetry.

Pulsed current-induced magnetization switching loops with $\beta$ = 1.35 nm (laser power = 14 mW) under various in-plane magnetic fields $H_x$ are shown in Figure 4c. Switching probabilities, defined as the ratio of $J_x$-induced to external field-induced $R_H^{max}$, of 16.4%, 63.6%, and 90.8% were obtained for $H_x$ = -200 Oe, $H_x$ = 0, and $H_x$ = 200 Oe, respectively. Note that the observed unequal switching between $H_x$ = -200 Oe and $H_x$ = 200 Oe is quite different from the case in conventional SOT-induced magnetization switching. By comparing the switching probabilities for $H_x$ = ±200 Oe and $H_x$ = 0 as functions of the laser power, as shown in Figure 4d, we ascribe a joint effect of two independent mechanisms for nonzero-$H_x$-assisted current-induced magnetization switching: the conventional SOT from the perpendicular asymmetry, and the SOT from the localized laser annealing induced lateral Pt-Co asymmetry. Particularly, the zero-field switching probability gradually increases from 0% to 90.2% (regarded as an approximately full switching due to the considerable current shunting effect in a Hall bar device without pillars[33], see the switching loop in Supplementary S4) on increasing the laser power from 0 mW to 16 mW, owing to an increasing $H_z^{eff}$ from the lateral Pt-Co asymmetry-induced SOT. Meanwhile, the anticlockwise switching with $H_x$ = -200 Oe in the as-deposited sample converts into clockwise switching for samples with 10 mW and higher laser powers, indicating a competition between two $H_z^{eff}$ with different signs. In contrast, the switching probability with $H_x$ = 200 Oe, which is only 11.9% for the as-deposited sample, increases rapidly to nearly full switching (88.6%) for 10 mW laser annealed samples (and maintains full switching for higher laser powers with a gradually decreasing $J_x^c$, as illustrated in Supplementary S5), indicating a constructive effect of two $H_z^{eff}$ with the same sign. These results are consistent with the interplay of effective fields: the sign of conventional SOT-induced $H_z^{eff}$ changes on reversing $H_x$, while the $β$-induced $H_z^{eff}$ stays constant under different $H_x$. This further illustrates the SOT from lateral Pt-Co asymmetry as an important effect, independent of the external out-of-plane spin current, for deterministically switching a PMA-FM layer.

The technology of local laser annealing is promising for fabricating scalable spintronic devices. A similar case is the matured proposal of heat-assisted magnetic recording (HAMR)[34], where a locally laser induced thermal effect assists the magnetization switching of independent magnetic recording units within tens of nanometers, which is under intensive production development now. Moreover, the fabrication of a lateral Pt-Co asymmetry may not be limited to our localized laser annealing method. Looking to the future, technologies capable of generating analogous lateral asymmetry inside a FM are expected to facilitate the current-induced magnetization switching without external out-of-plane spin current.

In summary, by introducing a lateral Pt gradient within the magnetic layer by localized laser annealing, an in-plane current-induced perpendicular effective magnetic field is generated which can deterministically switch the perpendicular magnetization at zero external magnetic field. The polarization and magnitude of the new SOT under an in-plane current depend only on the direction and degree of the lateral Pt-Co asymmetry, which are independent of the spin currents from neighboring layers or interfaces. The described lateral SOT offers a pathway to all electrical manipulation of spins without external out-of-plane spin current injection, and provides further insights into the in-plane current-induced SOTs in such asymmetric structures.

**Experimental Section**

*Thin film preparation:* The films were deposited at room temperature onto 0.5mm-thick Si wafers with a 190nm-thick thermal $SiO_2$ surface. D.C. magnetron sputtering was used to deposit the Pt and the Co layers. The base pressure of the chamber was less than $2\times10^{-6}$ Pa, and Ar gas was used for sputtering. The pressure of the chamber was $1.064\times10^{-1}$ Pa during deposition. No magnetic field was applied during the sputtering. The deposition rates for Pt and Co layers were controlled to be ~0.023 nm s$^{-1}$ and ~0.012 nm s$^{-1}$, respectively.

*Device fabrication:* Two types of Hall bar devices were used in this work. For Hall bar devices with pillars, as shown in Figure 1, two steps of standard electron-beam lithography (EBL) and Ar ion etching were used. First, the Pt(3 nm)/Co(0.5 nm)/Pt(2 nm) film was patterned into a Hall bar with channel width of 6 μm. Then, the central areas of the Hall bar crosses were patterned into 2×2 μm$^2$ square pillars, leaving the rest of the Hall bar with only the bottom Pt(3 nm) layer; For Hall bar devices without pillars, used in Figure 2 and Figure 4, Pt(3 nm)/Co(0.5 nm)/Pt($t_{Pt}$) films were used and all the three layers of each stack were patterned into a Hall bar with channel width of 6 μm.

*Localized laser annealing:* A laser with wavelength of 532 nm and power from 6 to 16 mW was used to locally anneal each sample in air atmosphere by sweeping along the *x*-direction with a velocity of 20 nm/s. The sweeping was realized by fixing the laser spot position meanwhile controlling the position and movement of the sample using a three-dimensional automated stage and a Thorlabs APT piezo controller, with an in-plane resolution of 5 nm. The locations of the localized laser annealing tracks were controlled at either the -*y* or the +*y* side of the pillar (Figure 1) or the cross area of the Hall bar (Figures 2 and 4).

*Measurement and characterization:* The current-induced magnetization switching and anomalous Hall effect measurements were carried out at room temperature with a Keithley 2602B as the current source and Keithley 2182 as the nanovoltmeter. The harmonic voltage measurements were conducted using two SR830 DSP lock-in amplifiers. The sample for STEM characterization was prepared using a Zeiss Auriga FIB system. The HAADF Z-contrast images, the BF images, and the EDS mappings were acquired on a spherical aberration-corrected FEI Titan Cubed Themis 60-300 operated at 200 kV.

**Supporting Information**

Supplementary Information is available from the Wiley Online Library or from the authors.

**Acknowledgements:**

This work was supported by National Key R&D Program of China (Grant No. 2017YFB0405700), by the NSFC (Grant No. 11474272 and 61774144), by Chinese Academy of Sciences (Grant No. QYZDY-SSW-JSC020, XDPB12, and XDB28000000), and by Beijing Natural Science Foundation Key Program (Grant No. Z190007).

**Competing Interests**

The authors declare no competing financial interests.

**Additional Information**

Correspondence and requests for materials should be addressed to K.W.

Supplementary Information

# Deterministic magnetization switching using lateral spin-orbit torque

*Yu Sheng[†], Yi Cao[†], Kevin William Edmonds, Yang Ji, Houzhi Zheng, and Kaiyou Wang[*]*

Y. Sheng, Dr. Y. Cao, Prof. Y. Ji, Prof. H. Zheng, Prof. K. Wang
State Key Laboratory for Superlattices and Microstructures
Institute of Semiconductors
Chinese Academy of Sciences
Beijing 100083, China
E-mail: kywang@semi.ac.cn

Dr. Y. Cao, Prof. K. Wang
Beijing Academy of Quantum Information Sciences
Beijing 100193, China

Prof. K. W. Edmonds
School of Physics and Astronomy
University of Nottingham
Nottingham NG7 2RD, United Kingdom

Prof. Y. Ji, Prof. H. Zheng, Prof. K. Wang
Center of Materials Science and Optoelectronic Engineering
University of Chinese Academy of Science
Beijing 100049, China

Prof. K. Wang
Center for Excellence in Topological Quantum Computation
University of Chinese Academy of Science
Beijing 100049, China

[†]These authors contributed equally to this work.

[*]Corresponding e-mail: kywang@semi.ac.cn

**Contents**

## S1. Magnetic properties of the Pt(3 nm)/Co(0.5 nm)/Pt($t_{Pt}$) stacks before and after localized laser annealing

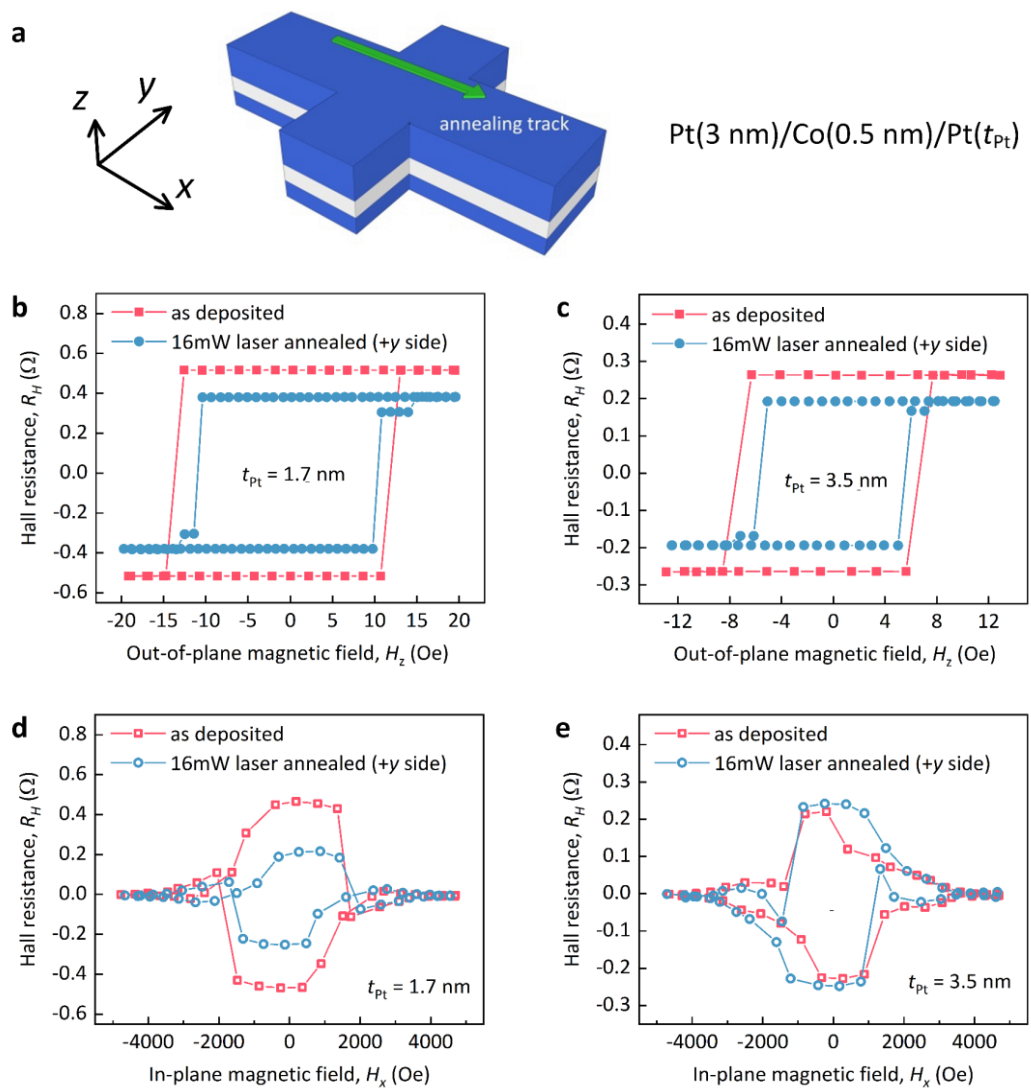

**Figure S1 | Schematic drawing of the laser annealing location and magnetic property changes of the stacks. a**, Schematic drawing of the laser track location for a Hall bar device. **b-e**, (**b-c**) Out-of-plane magnetic field-induced $R_H$-$H_z$ loops and (**d-e**) in-plane magnetic field-induced $R_H$-$H_x$ loops for the as deposited (red squares) and the +y side 16 mW locally laser annealed (blue circles) samples with stack structures of (**b, d**) Pt(3 nm)/Co(0.5 nm)/Pt(1.7 nm) and (**c, e**) Pt(3 nm)/Co(0.5 nm)/Pt(3.5 nm),

respectively.

All the stack structures used in this article are Si/SiO$_2$ substrate/Pt(3 nm)/Co(0.5 nm)/Pt($t_{Pt}$), where the top Pt layer $t_{Pt}$ ranges from 1.7 nm to 3.5 nm. By measuring the $R_H$-$H_z$ loops for samples with $t_{Pt}$ = 1.7 nm and $t_{Pt}$ = 3.5 nm, for representatives, the magnetic properties before and after the 16 mW localized laser annealing were examined. As shown in Figure S1b and S1c, both as deposited samples exhibit rectangle out-of-plane magnetic hysteresis loop, meanwhile, the hysteresis loops for 16 mW locally laser annealed samples keep good rectangularity but appear small jumps before reaching full magnetization switching. The small jumps in the hysteresis loops are supposed to be caused by the annealing track-induced pinning effect to the domain wall motions in the hall crosses. Furthermore, the magnetic anisotropy field $H_k$ (i.e. the in-plane magnetic field required for rotating a PMA moment to in-plane direction) were determined to be around 3500 Oe for both samples whenever before or after the localized laser annealing, as shown by the in-plane magnetic field-induced $R_H$-$H_x$ loops in Figure S1d and S1e, suggesting good PMA with roughly the same $H_k$ for all the samples studied in this article, regardless of the top Pt thicknesses and the localized laser annealing.

Note that there is a reduction of the maximum Hall resistance $R_H^{max}$ for both samples after the localized laser annealing. With the annealing induced relative displacement between Co and top Pt layers, changes of perpendicular saturation magnetization (which can be diminished to zero, when all the Co atoms moves above the top Pt layer and get oxidized) and/or interface condition in the laser annealed region are supposed to be responsible for this variation of anomalous Hall effect[1]. Following this way, the monotonically decreasing $R_H^{max}$ shown in Figure 4b can be interpreted as a continuous enhancement of the lateral Pt-Co asymmetry in the device with the increasement of laser power.

## S2. Harmonic measurement of the in-plane effective magnetic field

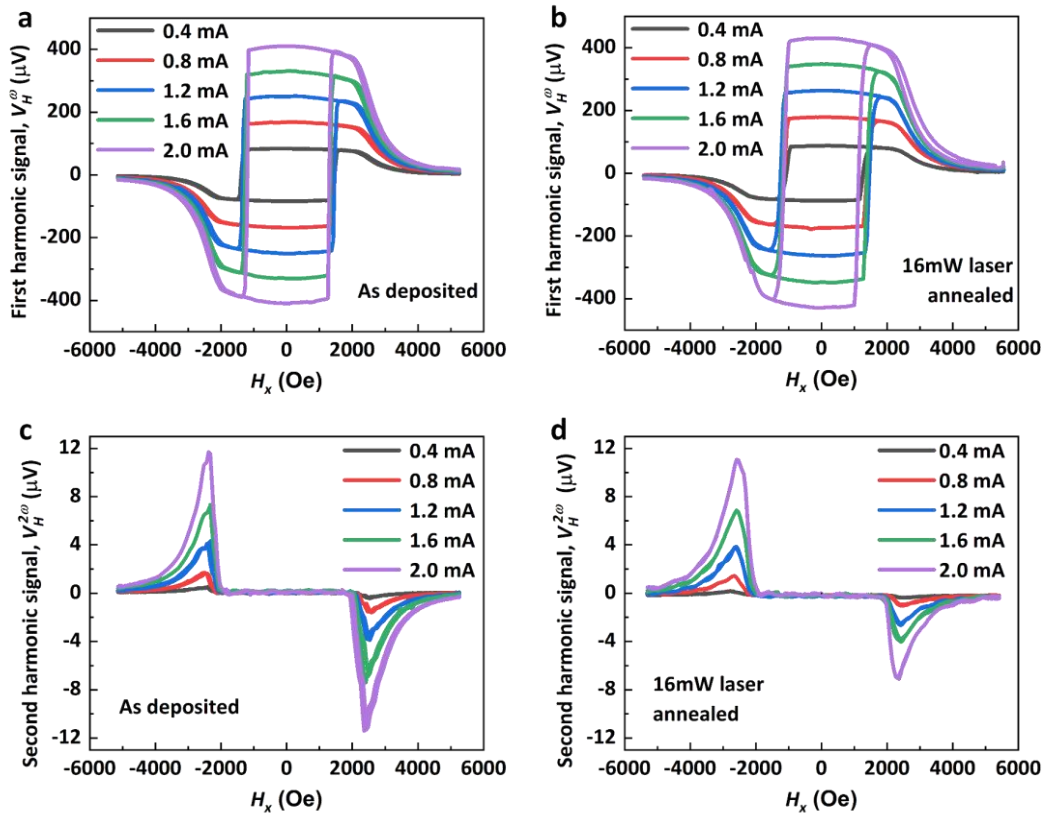

**Figure S2 | First (a, b) and second (c, d) harmonic measurement of the as deposited (a, c) and the +y side 16 mW locally annealed (b, d) Pt(3**

nm)/Co(0.5 nm)/Pt(2.6 nm) Hall bars (without pillars).** A.C. current with magnitudes from 0.4 to 2.0 mA were applied to the channel along $x$ direction, while first and second order Hall voltages were measured across the Hall arm in $y$ direction. The scanning in-plane magnetic field was applied parallel to the $x$ direction. The harmonic voltage measurements were conducted using two SR830 DSP lock-in amplifiers.

Harmonic measurements provide a powerful probe of the current-induced in-plane effective fields from the damping-like and the field-like torques. As illustrated in our previous work (manuscript Ref. 12), pillared Pt/Co/Pt device gives negligible field-like torque but significant damping-like torque with obviously tilted second harmonic Hall curve near $H_x = 0$. In the case of the as deposited Pt(3 nm)/Co(0.5 nm)/Pt(2.6 nm) Hall bars without pillar, however, almost flat and unchanged second harmonic signal curves were observed near $H_x = 0$ for the as-deposited and the 16 mW laser annealed samples with various applying a.c. channel current, as shown in Figure S2c and S2d. Although the peaks shape of second harmonic Hall signals show there are damping-like torque with same directions before and after the localized laser annealing[2], but the flat slopes of them near $H_x = 0$ suggest negligible small values of the damping-like torques. This result agrees with the largely cancelled out-of-plane spin current injections from the bottom by the top Pt layers, but the value of which is small and remained nearly unchanged after localized laser annealing.

## S3. HAADF and BF images of the locally laser annealed Pt/Co/Pt sample

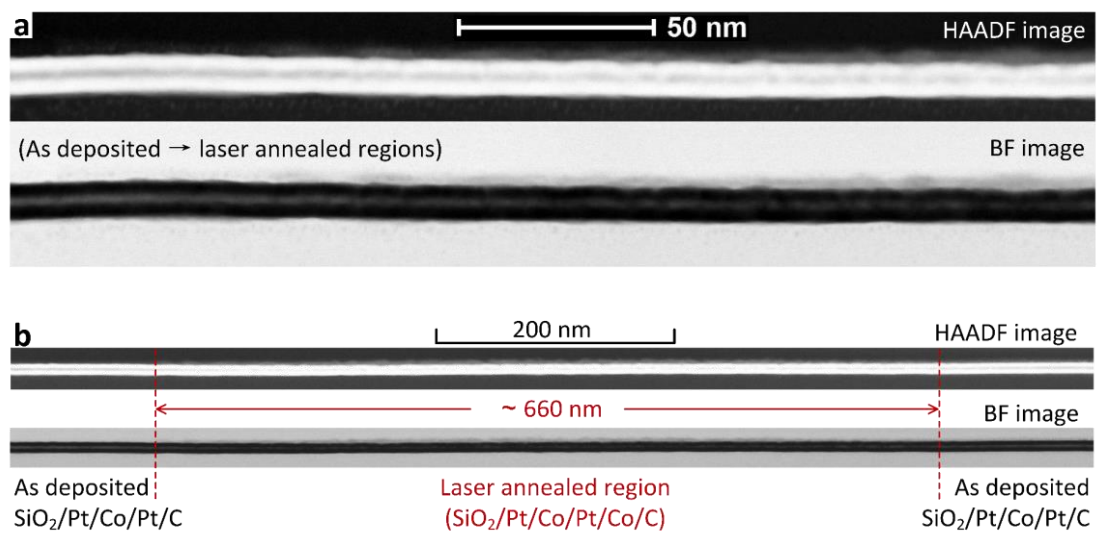

**Figure S3 | HAADF and BF images of the 12 mW locally laser annealed SiO₂ substrate surface/Pt(3 nm)/Co(3 nm)/Pt(3 nm)/C capping layer STEM sample.** Brighter contrast represents heavier (Pt) / lighter (Co, Si, and C) elements in the HAADF / the BF images, respectively. **a**, HAADF and BF images of the gradient area from the left as deposited to the right laser annealed regions. **b**, Extended HAADF and BF images covering the whole 12 mW laser annealed region. The boundaries between the as deposited and the laser annealed regions are defined at positions where distinguishable Co-Pt intermigration appears. Magnify the figure to see details.

The contrast in a BF image is inverse to a HAADF image, thus providing a complementary view to the macrostructure. A comparison in the BF and the HAADF images of the same area with Figure 3a is shown in Figure S3a, where we can find that as deposited middle Co layer and the laser annealing-induced migrated top Co layer are more obvious in the HAADF image and the BF image, respectively.

Figure S3b shows extended HAADF and BF images covering the whole laser annealed region. By defining the boundary of the as deposited and the laser annealed regions at the position where distinguished Co-Pt intermigration appears, the total width of the 12 mW laser annealed region was obtained as around 660 nm.

### S4. Zero-field $R_H$-$J_x$ loop for the 16 mW locally laser annealed sample

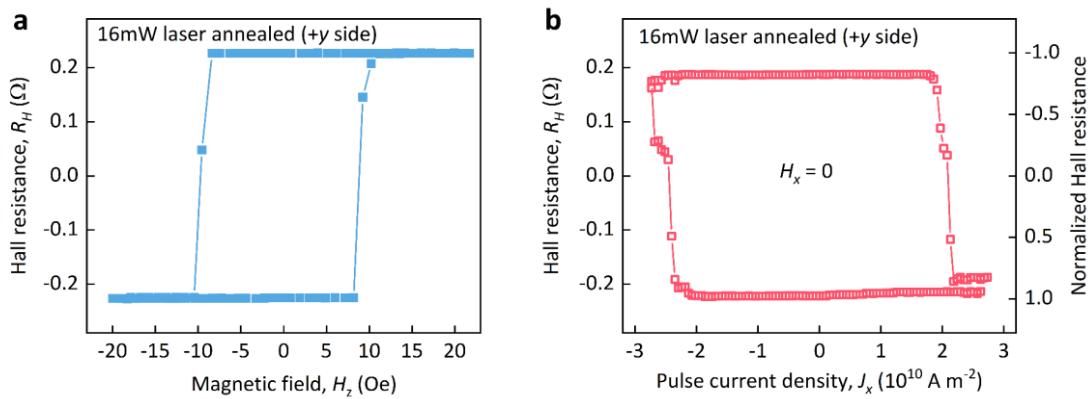

**Figure S4 | The magnetic field-induced and the zero-field current-induced magnetization switching after localized laser annealing. a-b,**

(**a**) The $R_H$-$H_z$ loop and (**b**) the zero-field $R_H$-$J_x$ loop for the +y side 16 mW locally laser annealed Pt(3 nm)/Co(0.5 nm)/Pt(2.6 nm) Hall bar. The initial magnetic state was set fully downwards by $H_z$ = -200 Oe before the current-induced magnetization switching measurement.

The out-of-plane magnetic field-induced as well as the zero-field current-induced magnetization switching loop of the +y side 16 mW locally laser annealed Pt(3 nm)/Co(0.5 nm)/Pt(2.6 nm) Hall bar device are shown in Figure S4a and Figure S4b, respectively. Considering the unneglectable current shunting effect in a Hall bar device without pillars, the measured Hall resistance $R_H$ is not only contributed by the central cross area of the Hall bar, but also partly from the arms. Therefore, even a full magnetization switching is realized in the cross area, it is not possible for the measured $R_H$ to reach the $H_z$-induced maximum Hall resistance $R_H^{max}$. In this way, we regard the switching probability of ~90% in Figure 4d and Figure S4b as an approximately full current-induced magnetization switching.

## S5. $J_x^c$ under $H_x$ = 200 Oe for laser powers above 10 mW

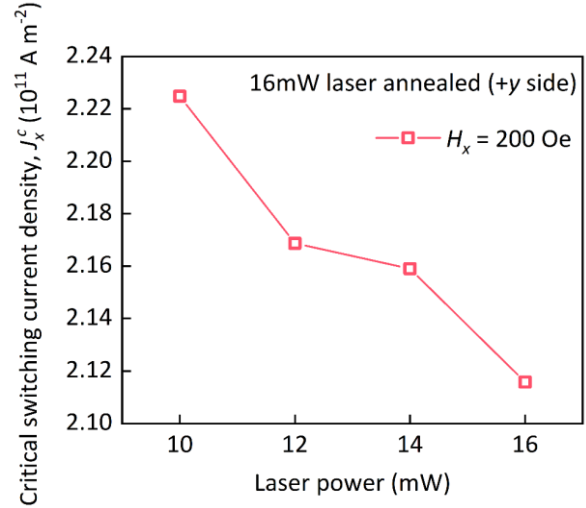

**Figure S5 | In-plane magnetic field-assisted critical switching current density as a function of the laser power.** The samples are Pt(3 nm)/Co(0.5 nm)/Pt(2.6 nm) Hall bar devices with localized laser annealing at the +*y* side. The assistant in-plane magnetic field $H_x$ is fixed at 200 Oe and the laser powers ranges from 10 mW to 16 mW.

The critical current density $J_x^c$ in this article is defined as the average values of $J_x$ for the magnetization switching from +*z* to -*z* directions[3]. The $J_x^c$ is usually used to evaluate the level of difficulty for a current-induced magnetization switching, however, for the cases in Figure 4d, where most of the switching probabilities are below 50%, the switching probability should be a more important indicator rather than the $J_x^c$.

For laser powers above 10 mW in Figure 4d, whose current-induced magnetization switching probabilities are all around 90% with the assistance of $H_x$ = 200 Oe, their $J_x^c$ were then considered for estimating the switching difficulties. As shown in Figure S5, the $J_x^c$ slightly decreases as the laser power increases from 10 mW to 16 mW, corresponding to enhanced current-induced perpendicular effective magnetic fields $H_z^{eff}$ due to stronger lateral Pt-Co asymmetries with the growing laser power.